\documentclass{article}

\usepackage{times}

\usepackage{graphicx} 
\usepackage{subcaption}
\usepackage{amsmath}

\usepackage{enumitem}

\usepackage[sort&compress,numbers,square,comma,authoryear]{natbib}

\usepackage{algorithm}
\usepackage{algorithmic}

\usepackage{hyperref}

\usepackage[accepted]{cs541}
\setcitestyle{sort&compress,numbers,square,comma}

\icmltitlerunning{Enhanced Residual Networks for Context-based Image Outpainting}

\begin{document} 

\twocolumn[
\icmltitle{Enhanced Residual Networks for Context-based Image Outpainting}



\icmlsetsymbol{equal}{*}

\begin{icmlauthorlist}
\icmlauthor{Przemek Gardias}{wpi}
\icmlauthor{Eric Arthur}{wpi}
\icmlauthor{Huaming Sun}{wpi}
\end{icmlauthorlist}

\icmlaffiliation{wpi}{Worcester Polytechnic Institute}

\icmlcorrespondingauthor{}{\{pmgardias, etarthur, hsun2\}@wpi.edu}

\icmlkeywords{image outpainting, generative adversarial network, context encoder, computer vision, deep learning, machine learning, ICML}

\vskip 0.3in
]



\printAffiliationsAndNotice{}  

\begin{abstract} 
Although humans perform well at predicting what exists beyond the boundaries of an image, deep models struggle to understand context and extrapolation through retained information. This task is known as image outpainting and involves generating realistic expansions of an image’s boundaries. Current models use generative adversarial networks to generate results which lack localized image feature consistency and appear fake. We propose two methods to improve this issue: the use of a local and global discriminator, and the addition of residual blocks within the encoding section of the network. Comparisons of our model and the baseline’s L1 loss, mean squared error (MSE) loss, and qualitative differences reveal our model is able to naturally extend object boundaries and produce more internally consistent images compared to current methods but produces lower fidelity images.\footnote[2]{A repository containing relevant source code, and accompanying instructions, is accessible via \href{https://github.com/etarthur/Outpainting}{GitHub}.}
\end{abstract} 

\section{Introduction}
The use of generative adversarial models has led to many new developments in the computer vision field. The task known as image inpainting, which aims to enhance or restore quality to parts of an image, requires a model to retain the context of the image while completing missing regions of the image, and iteratively improve the process of the generations. The problem explored in this paper is a significantly more ambiguous task than its inpainting alternative, and is commonly known as image outpainting. For this task, the input image boundaries are extrapolated outside of the original content of the input based upon the context of the image \citep{sabini_painting_2018}.

\begin{figure}
	\captionsetup[subfigure]{labelformat=empty}
    \centering
    \begin{subfigure}[b]{0.15\textwidth}
        \includegraphics[width=\textwidth]{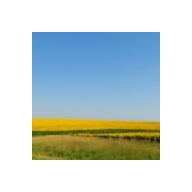}
        \caption{Masked input}
    \end{subfigure}
    \hfill
    \begin{subfigure}[b]{0.15\textwidth}
        \includegraphics[width=\textwidth]{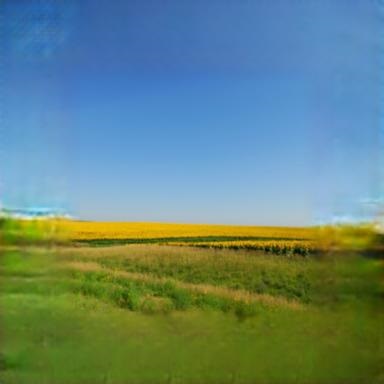}
        \caption{Output}
    \end{subfigure}
    \hfill
    \begin{subfigure}[b]{0.15\textwidth}
        \includegraphics[width=\textwidth]{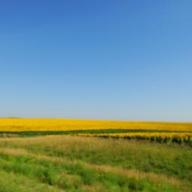}
        \caption{GT}
    \end{subfigure}
  	\caption{Example of image outpainting.}
  	\label{fig:fig1}
\end{figure}

Image outpainting, sometimes referred to as image context interpretation and extrapolation, requires the context of the image to be retained by the encoder to accurately re-generate the known input image, then create new portions to append to the edges of the image as shown in Figure \ref{fig:fig1}. There exists research on this task using generative adversarial networks, however one of the biggest issues with image outpainting is the new additions to the image, known as hallucinations, typically appear lower quality than the base image and therefore can easily be distinguished as fake by the human eye \citep{sabini_painting_2018}. Therefore, to effectively create higher quality hallucinations, better image evaluation methods are required to match the discrimination abilities of human interpretation.

\subsection{Research Contributions}
We provide insight into the use of a local discriminator alongside a global discriminator to improve the quality of hallucinations. Specifically, we analyze how dual discriminators assist in smoothing the sudden change in quality between hallucinations and original image. Additionally, we successfully implement residual blocks applied to the context encoder portion of our generative network to improve the quality of images. We evaluate the performance of these two networks by comparing to the baseline model provided by Van Hoorick using a time based remedy to the generator loss function \citep{van_hoorick_image_2020}.

\section{Related Work}
In this section, we briefly review the previous works relating to this paper in three sub-fields: Generative Adversarial Networks, Image Inpainting, and Image Outpainting.

\subsection{Generative Adversarial Networks}
Deep generative models have shown success in various tasks of image and video generation problems. By training the generator and discriminator in tandem, the generator can capture the real data distribution and create more realistic images given the latent input. However, GANs are often prone to collapse and are difficult to train. Therefore, some newer techniques have been applied attempting to reduce this issue such as the addition of residual blocks to the encoder component of the generator specifically researched within a super-resolution task context \citep{lim_enhanced_2017}.

\subsection{Image Inpainting}
Image inpainting is the task of recovering missing regions or enhancing regions within images. Pathak et al. demonstrate that context encoder can achieve realistic results for generation of novel images sections for inpainting purposes \citep{pathak_context_2016}. Recent improvements include a local discriminator to improve the generated feature consistency within the missing sections of the input image by identifying poor generations only within the area of the inpainted sections \citep{iizuka_globally_2017}. More recently, Liu et al. apply partial convolutions to improve overall consistency of generated images \citep{liu_image_2018}.

\subsection{Image Outpaint}
Image outpainting refers to predicting the region beyond the borders of an image. Compared to image inpainting, it has been significantly less studied. Although there are promising works on this problem, they include limitations such as application to very specific domain datasets or result in low quality of hallucinations \citep{van_hoorick_image_2020}\citep{wang_wide-context_2019}. More recent work utilizes a deep neural network with both global and local discriminator, recurrent content transfer, skip horizontal connection, and a global residual block to achieve impressive results \citep{yang_very_2019}. However, this approach focuses  on recursive horizontal hallucinations on a natural scenes dataset, resulting in a network which prioritizes clearly defined horizontal lines often found in landscape images.

\section{Proposed Method}
In this section, we provide an overview of our architecture. Since the task is generative we use a GAN in a similar context-encoding method compared to current methods \citep{van_hoorick_image_2020}\citep{pathak_context_2016}. Our GAN architecture contains two major additions: the first being the use of residual blocks in the generator, and the second being the use of two discriminators. Below, we explore our architecture choices and some design decisions behind it.

\begin{table}[h] 
  \centering  
    \begin{tabular}{lcc}  
    \hline
    \textbf{layer} & \textbf{output size} & \textbf{parameters}\\ 
    \hline \hline
      Conv & 16$\times$192$\times$192 & 5$\times$5, stride=1 \\
    \hline
      ResBlock & 128$\times$96$\times$96 & 3$\times$3, stride=2 \\
    \hline
      ResBlock & 256$\times$48$\times$48 & 3$\times$3, stride=2 \\
    \hline
      ResBlock & 256$\times$48$\times$48 & 3$\times$3, stride=1 \\
    \hline
      ResBlock & 256$\times$48$\times$48 & 3$\times$3, stride=1 \\
    \hline
      ResBlock & 256$\times$48$\times$48 & 3$\times$3, stride=1 \\
    \hline
      Conv & 256$\times$48$\times$48 & 3$\times$3, stride=1 \\
    \hline
      ReLU & 256$\times$48$\times$48 & None \\
    \hline
      Conv & 256$\times$48$\times$48 & 3$\times$3, stride=1 \\
    \hline
      ReLU & 256$\times$48$\times$48 & None \\
    \hline
      Conv & 256$\times$48$\times$48 & 3$\times$3, stride=1 \\
    \hline
      ReLU & 256$\times$48$\times$48 & None \\
    \hline
      Trans-Conv & 128$\times$96$\times$96 & 4$\times$4, stride=1 \\
    \hline
      ReLU & 128$\times$96$\times$96 & None \\
    \hline
      Conv & 128$\times$96$\times$96 & 3$\times$3, stride=1 \\
    \hline
      ReLU & 128$\times$96$\times$96 & None \\
    \hline
      Trans-Conv & 64$\times$192$\times$192 & None \\
    \hline
      ReLU & 64$\times$192$\times$192 & None \\
    \hline
      Conv & 32$\times$192$\times$192 & 3$\times$3, stride=1 \\
    \hline
      ReLU & 32$\times$192$\times$192 & None \\
    \hline
      Conv & 3$\times$192$\times$192 & 3$\times$3, stride=2 \\
    \hline
      Sigmoid & 3$\times$192$\times$192 & None \\
    \hline
    \end{tabular}
  
  \caption{Architecture of generator $G$, including parameters. Trans-Conv is transposed convolution.} 
  \label{tab:1} 
\end{table}

\begin{table}[h] 
  \centering  
    \begin{tabular}{lc}  
    \hline
    \textbf{Layer} & \textbf{Parameters}\\ 
    \hline \hline
      Conv & 3$\times$3, stride=$s_{in}$ \\
    \hline
      ReLU & None \\
    \hline
      Conv & 3$\times$3, stride=$s_{in}$ \\
    \hline
      ReLU & None \\
    \hline
      Add & None \\
    \hline
    \end{tabular}
  
  \caption{Architecture of residual block used in generator $G$, including parameters. The residual block applies input strides $s_{in}$ to the convolution layer. Add is an element-wise addition of block input tensor $x$ and output tensor of previous layer.} 
  \label{tab:2}
\end{table}

\subsection{Generator}
We propose a GAN architecture that aims to encode the context of the image and use deconvolutional layers to then generate the resulting image while retaining feature consistency. We apply a residual model to the generative network, first convolving the image down into its feature space, then using deconvolution to upsample the image into its expanded version. Although typical models that contain residual blocks often include batch normalization within the blocks, we do not include this activation layer per demonstrations of improvement in the deblurring work of Nah et al \citep{nah_deep_2018}. The encoder section of the generative network contains multiple residual blocks in an attempt to improve image quality. Table \ref{tab:1} outlines the layers of our generative network, including both the encoder and decoder components. Table \ref{tab:2} showcases the individual layers which are used to construct a residual block.

\subsection{Discriminator}
Our discriminator network $D$ consists of dual discriminators. One discriminator $D_G$ takes as input the entire generated image and applies a convolutional network to identify the image as real or generated, while the second discriminator $D_L$ is local, and is restricted solely to the boundaries between the input image and the hallucinations. These discriminator outputs are averaged as such, where $\epsilon$ represents the binary mask which is originally applied to the training images: 
\begin{align}
	D(x) &= \frac{D_L(\epsilon \times x) + D_G(x)}{2}
\end{align}

The purpose of the local discriminator is to attempt to remove the clearly fake hallucinations that result from low-quality generations or manifestations, which immediately appear fake to the human eye. Since these happen in the boundary between the input image and the hallucinations, the local discriminator focuses on the hallucination alone.

\begin{table}[h] 
  \centering  
    \begin{tabular}{lc}  
    \hline
    \textbf{Layer} & \textbf{Parameters}\\ 
    \hline \hline
      Conv & 3$\times$3, stride=2 \\
    \hline
      InstanceNorm & None \\
    \hline
      LeakyReLU & negative slope=0.2 \\
    \hline
      Conv & 3$\times$3, stride=2 \\
    \hline
      InstanceNorm & None \\
    \hline
      LeakyReLU & negative slope=0.2 \\
    \hline
      Conv & 3$\times$3, stride=2 \\
    \hline
      InstanceNorm & None \\
    \hline
      LeakyReLU & negative slope=0.2 \\
    \hline
      Conv & 3$\times$3, stride=2 \\
    \hline
      InstanceNorm & None \\
    \hline
      LeakyReLU & negative slope=0.2 \\
    \hline
      Conv & 3$\times$3, stride=1 \\
    \hline
      InstanceNorm & None \\
    \hline
      LeakyReLU & negative slope=0.2 \\
    \hline
      Conv & 3$\times$3, stride=1 \\
    \hline
    \end{tabular}
  
  \caption{Architecture of global and local discriminators $D_G$ and $D_L$, respectively.} 
  \label{tab:3}
\end{table}

\begin{figure*}
	\captionsetup[subfigure]{labelformat=empty}
    \centering
    \begin{subfigure}[b]{0.175\textwidth}
        \includegraphics[width=\textwidth]{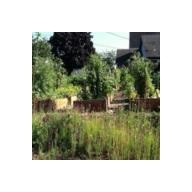}
        \caption{Masked input}
    \end{subfigure}
    \hfill
    \begin{subfigure}[b]{0.175\textwidth}
        \includegraphics[width=\textwidth]{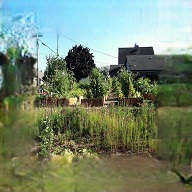}
        \caption{Baseline}
    \end{subfigure}
    \hfill
    \begin{subfigure}[b]{0.175\textwidth}
        \includegraphics[width=\textwidth]{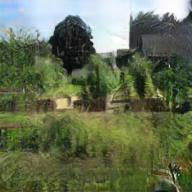}
        \caption{Local}
    \end{subfigure}
    \hfill
    \begin{subfigure}[b]{0.175\textwidth}
        \includegraphics[width=\textwidth]{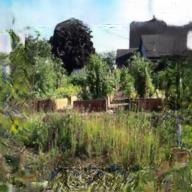}
        \caption{Residual}
    \end{subfigure}
    \hfill
    \begin{subfigure}[b]{0.175\textwidth}
        \includegraphics[width=\textwidth]{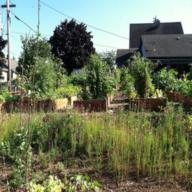}
        \caption{GT}
    \end{subfigure}
  	\caption{Comparison of results for different models. Note the baseline result appears larger but contains the same ratio of hallucination to input pixels as the other generated images.}
  	\label{fig:fig2}
\end{figure*}

\begin{figure}
	\captionsetup[subfigure]{labelformat=empty}
    \centering
    \begin{subfigure}[b]{0.15\textwidth}
        \includegraphics[width=\textwidth]{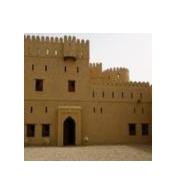}
        \caption{Masked input}
    \end{subfigure}
    \hfill
    \begin{subfigure}[b]{0.15\textwidth}
        \includegraphics[width=\textwidth]{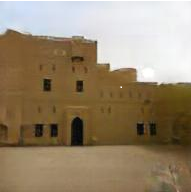}
        \caption{Local}
    \end{subfigure}
    \hfill
    \begin{subfigure}[b]{0.15\textwidth}
        \includegraphics[width=\textwidth]{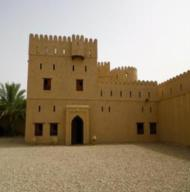}
        \caption{GT}
    \end{subfigure}
  	\caption{Outpainting example of the local discriminator model.}
  	\label{fig:fig3}
\end{figure}

\section{Experiment}
To evaluate the effectiveness of our model we compare MSE and L1 loss between each of our models. Furthermore, we qualitatively analyze the outputs from our models due to the inherently visual nature of this problem.

We use MIT CSAIL Places365-Standard, a large scale scene dataset used to train models for image context and recognition \citep{zhou_places_2017}. This dataset contains around 2 million images, combined, of random scenes from outdoor and indoor scenery, including both simple landscapes and detailed, object-heavy images. We use this dataset as our primary method of evaluation because of its usage in Van Hoorick \citep{van_hoorick_image_2020}. Therefore, a direct comparison of MSE loss is more meaningful. Unfortunately, due to computing time constraints, training on the whole of the dataset was deemed implausible, therefore we drastically reduced the size of the training set to approximately 25,000 images and trained for 50 epochs. The size of the validation set was reduced to be relatively proportional to the training set decrease. During this process, we kept images which were chosen randomly from the full dataset, a process which adds some inherent weaknesses to the training of our model. To take into consideration these limitations, we trained three models: 
\begin{itemize}
\setlength{\partopsep}{0pt}
\setlength{\topsep}{0pt}
\setlength{\itemsep}{0pt}
	\item{The model proposed by Van Hoorick, which is herein referred to as ``baseline''.}
	\item{A model which adds on a local discriminator to the existing context encoder architecture, averaging the results of the dual discriminators. This model is herein referred to as ``local''.}
	\item{A model which uses residual blocks as part of the context encoder network, as well as includes the dual discriminators. This model is herein referred to as ``residual''.}
\end{itemize}
To address the weakness which training on a subset of the dataset introduce, we train the baseline model on the same training set which is used for the latter two models.

We trained each of our models with a fixed learning rate of $\alpha=0.0003$ and two Adam optimizers with $\beta_1=0.5, \beta_2=0.999$, following the same training considerations proposed by Van Hoorick \citep{van_hoorick_image_2020}. The loss functions are as follows:
\begin{align}
  L_{rec} &= ||x-G(x)||_1 \\
  L_{adv} &= ||D(G(x))-1||_2^2 \\
  L_G &= \lambda_{rec}L_{rec} + \lambda_{adv}L_{adv} \\
  L_D &= ||D(x)-1||_2^2 + ||D(G(x))-0||_2^2
\end{align}
We apply a time based remedy to the generator loss function that punishes the generator heavier for bad outputs as time progresses:
\begin{align}
  \lambda_{adv}(n) = \begin{cases}
    0.001, & \text{if }n\leq10 \\
    0.005, & \text{if }10<n\leq30 \\
    0.015, & \text{otherwise}
  \end{cases}
\end{align}
Since humans still perform vastly greater than deep learning networks at this task we conclude our experiment with a qualitative discussion about the quality of hallucinations. As many previously mentioned models in this domain create noise or extrapolations in the hallucinations that clearly appear faulty, we aim to manually evaluate whether our model rectifies this issue or these manifestations are still present.

\section{Results}
We first provide a comparison of the losses for each model in Table \ref{tab:4} that demonstrates the residual model performs marginally better with respect to L1 and adversarial loss. However, as a comparison of the loss values does not cover all of the results in detail, we provide a further discussion of the qualitative results, which provides insights to the strengths and weakness of our approaches.

\begin{table}[h] 
  \centering  
    \begin{tabular}{lrrr}
    \hline
    \textbf{Model} & \textbf{L1} & \textbf{MSE} & \textbf{Adversarial}\\ 
    \hline \hline
      Baseline & $0.0956$ & $0.047$ & $0.1278$ \\
      Local Discriminator & $0.897$ & $1.044$ & $0.1014$ \\
      Residual Encoder & $0.08$ & $0.7814$ & $0.0941$ \\
    \hline
    \end{tabular}
  
  \caption{Comparison of all loss results across each of the models.} 
  \label{tab:4}
\end{table}

The baseline model performs better at directly reconstructing the images compared to their ground truths, as demonstrated by the significantly lower MSE, however the local discriminator and residual models result in an L1 loss which is marginally lower than the baseline model's L1 loss. Since we set out not to improve the reconstruction but rather the quality of the reconstructed image, the change in MSE is expected. Finally,  moderately lower adversarial loss of the local and residual models reflects our goal with the implementation of dual discriminators, demonstrating their strength in determining fake generated images.

We aim to create a model that generates cleaner or higher quality hallucinations such that it is not evident to the human eye if it is fake or real. Per Figure \ref{fig:fig2}, we can demonstrate our approach is able to achieved a clearer continuous output. As shown below the second image is the output from the baseline model, and it is quite evident where the original image ends and the new hallucinations appear. Compared to the output from our local discriminator model, it is much harder to tell where the original image ends and the hallucinations appear. This is because our model is able to effectively continue object boundaries into the hallucinations which the baseline cannot. A side effect of this is that the whole image becomes blurrier which explains the increase in MSE, however qualitatively we believe the outputs show that our local discriminator model does a significantly better job at creating continuous realistic hallucinations compared to the baseline as we set out to achieve. Finally, our residual blocks model appears to revert the beneficial qualities of having two discriminators as shown by the obvious boundaries in the fourth image of figure \ref{fig:fig2}. The residual blocks allowed for the input image to remain high quality, but the use of a local discriminator lowered the quality of the hallucinations resulting in a model very close in effectiveness to the baseline.

\begin{figure}
	\captionsetup[subfigure]{labelformat=empty}
    \centering
    \begin{subfigure}[b]{0.25\textwidth}
        \includegraphics[width=\textwidth]{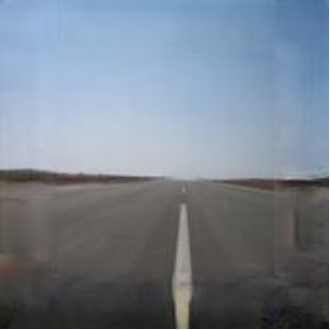}
    \end{subfigure}
  	\caption{Outpainting result example for the residual encoder model.}
  	\label{fig:fig4}
\end{figure}

\begin{figure}
	\captionsetup[subfigure]{labelformat=empty}
    \centering
    \begin{subfigure}[b]{0.25\textwidth}
        \includegraphics[width=\textwidth]{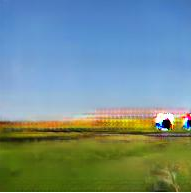}
    \end{subfigure}
  	\caption{Example of an anomaly within the hallucination, generated during the training process.}
  	\label{fig:fig5}
\end{figure}

With further examination we can report that the local discriminator model may reduce the quality of the input image, but by doing so can create more realistic outputs as shown in Figure \ref{fig:fig3}. The detail in the original image is not prioritized to be preserved but instead the overall continuity of the image is. For example, the individual peaks atop the structure are lost but the new generations match the quality of the blurred input image. Furthermore, it is evident that our model is able to effectively continue object boundaries into the hallucinations. Comparatively, the baseline model struggles to do this, and therefore our model provides a large improvement in this regard. Because of this the output image looks quite believable as there is internal consistency within the image that is lost in the baseline model.

Next, our residual blocks model which contains both a local and discriminator seemed to revert some of the benefits of having both discriminators. As shown in Figure \ref{fig:fig4}, the residual blocks model allowed for the input image to remain high quality but since it uses the local discriminator the hallucinations are of reduced quality. These two effects combined creates a clearly fake output image where the boundary of the original input image is clearly defined. The output appears marginally better than the baseline, however the hallucinations remain lower quality even though the entirety of the image is of higher quality than the local discriminator model's results.

Finally, due to the inclusion of the local discriminator our models place a significant emphasis on the hallucinations. This, combined with the fact that we do not use batch normalization, increases the likelihood of erroneous anomalies, as is exemplified in Figure \ref{fig:fig5}. Since our model attempts to blur the boundary zones between the base image and the hallucinations the anomaly is stretched and warped across the image.

\section{Discussion}
A major limitation of our experiment was the size of the training set and number of epochs we were able to fit in our limited computing time. Using just 200,000 images from the  Places365 dataset on the single NVIDIA Tesla P4 we used for training would have taken approximately a week for 200 epochs. Therefore, we reduced the dataset down to approximately 25,000 images randomly. The effects of this blind reduction of training data resulted in our models struggling to learn complex features such as humans and water details. This problem is even more compounded since humans were included in only a small subset of the images we used. This limitation could have been overcome with more time or better computational capabilities to train on a larger dataset. Alternatively, a smaller dataset could be used which requires significantly less information to be contained within the latent space. Furthermore, our reduced dataset could have been manually picked to only include images of landscapes to focus on one specific image type, albeit this approach would take an unreasonably excessive amount of tedious effort to properly 

Another limitation to our approach is that our models use a fixed average between the global and local discriminator. Therefore, it weighs the overall image and the boundary areas equally. This issues relates to the creation of anomalies, as shown in \ref{fig:fig5}, and why the images appear blurrier or lower quality overall. The easiest way to rectify this would be to include a method of back-propagate the error to the discriminator-combining weight and allow the models to learn how to combine the discriminator outputs. The models would then hopefully be able to balance higher quality images and consistency whereas our discussed approach focuses exclusively on consistency.

Since the MSE loss in this task is the measure of how well the models can reconstruct the ground truth image it may not have been an appropriate measurement for our specific problem of increasing the hallucination boundary quality of the images. Since our local discriminator and global discriminator have separate goals the MSE loss of our models was ultimately higher than the baseline. Therefore, the quantitative measure of MSE loss fails to reveal how our models excel. Unfortunately, we were unable to find another loss function for image consistency and retrain in time to publish these results.
	
Finally, the residual blocks model demonstrate increased image fidelity of the original input image inside of the hallucinations. Adding the residual blocks to the local discriminator model seems to ``revert'' the focus of the generator closer to the baseline, as the boundary regions become more defined. Therefore, we recommend further experimentation into the usage of residual blocks as part of the encoding network, including using residual blocks after the generation as a post-processing technique to construct higher quality output images.

\section{Conclusions and Future Work}
Image outpainting is an attractive task that currently suffers from low quality or obviously fake outputs. The models we created show that the use of a local and global discriminator allows for higher consistency within the output images and object boundaries may be continued into the newly generated areas. This is an improvement over other models and shows promise into ways to improve output quality. Further research is needed to improve the fidelity of these outputs, however the consistency of these images makes it significantly harder to detect fake outputs compared to other models.

Obvious improvements may be conducted related to the architecture and training of our models. Below we identify possible future work to extend our contribution to the problem:
\begin{itemize}
\setlength{\partopsep}{0pt}
\setlength{\topsep}{0pt}
\setlength{\itemsep}{1pt}
	\item{Comprehensive training on the entire two million image dataset for an increased number of epochs will yield significantly cleaner and robust results.}
	\item{The use of a local and global discriminator greatly improves object boundaries but dampens the overall quality of the image. This may be rectified by allowing the model to learn a combining weight for the combination of the dual discriminators using back-propagation.}
	\item{A more appropriate loss function to reinforce feature consistency across the boundary of the input image and the hallucination. For example, a loss function which rewards finer generated object boundaries, possibly even drawing from traditional computer vision techniques for feature detection across the ground truth and outpainting result.}
\end{itemize}

\bibliographystyle{unsrt}
\bibliography{paper}

\end{document}